\documentstyle[11pt]{article}

\let\tsection\section
\renewcommand{\section}{\setcounter{equation}{0}\tsection}

\def\F{{\cal F}}\def\Feq{{\cal F}_{\rm eq}}
\def\G{{\cal G}}
\let\cp\nu
\let\fp\lambda
\def\fncomma{$^{\rm ,}$}

\def\rhob{\bar\rho}

\begin{document}
\begin{center} 
LARGE DEVIATION OF THE DENSITY PROFILE IN THE STEADY STATE OF THE OPEN
  SYMMETRIC SIMPLE  EXCLUSION PROCESS

\vskip10pt

B. Derrida\footnote{Laboratoire de Physique Statistique,
Ecole Normale Sup\'erieure, 24 rue Lhomond, 75005 Paris, France; 
email derrida@lps.ens.fr.},
J. L. Lebowitz\footnote{Department of Mathematics,
Rutgers University, New Brunswick, NJ 08903; email lebowitz@math.rutgers.edu,
speer@math.rutgers.edu.}\fncomma\footnote{Also Department of Physics, 
Rutgers.},  and E. R. Speer${}^{\tiny 2}$

\end{center} 

\vskip20pt
\noindent {\bf Abstract}
We consider an open one dimensional lattice gas on sites $i=1,\dots,N$, with
particles jumping independently with rate $1$ to neighboring interior empty
sites, the {\it simple symmetric exclusion process}.  The particle
fluxes at the left and right boundaries, corresponding to exchanges with
reservoirs at different chemical potentials, create a stationary
nonequilibrium state (SNS) with
a steady flux of particles through the system.  The mean density
profile in this state, which is linear, describes the
typical behavior of a macroscopic system, i.e., this profile 
occurs with probability 1 when $N \to
\infty$.  The probability of microscopic configurations
corresponding to some other profile $\rho(x)$, $x = i/N$, 
has the asymptotic form 
$\exp[-N {\cal F}(\{\rho\})]$; $\cal F$ is the {\it large deviation
functional}.  In contrast to equilibrium systems, for which 
${\cal F}_{eq}(\{\rho\})$ is just
the integral of the appropriately normalized 
local free energy density,
the $\cal F$ we find here 
for the nonequilibrium system is a nonlocal function of
$\rho$.  This gives rise to the long range 
correlations in the SNS predicted by fluctuating hydrodynamics 
and suggests similar non-local
behavior of $\cal F$ in general SNS, where the long range correlations have
been observed experimentally.

\newpage

\section{Introduction\label{introduction}}

The extension of the central object of equilibrium
statistical mechanics, entropy or free energy, to stationary nonequilibrium
systems in which there is a
transport of matter or energy has proved difficult.
One knows from various approximate theories like fluctuating
hydrodynamics  that such systems exhibit long range correlations
very different from those of equilibrium systems \cite{Sch,DKS,HS1}.
These correlations extend over macroscopic distances,
as has been established rigorously in some models, and 
reflect the intrinsic nonadditivity of such systems. 
They  have been measured experimentally in a fluid with a steady
heat current \cite{DKS,Exp}. Their derivation from a well defined macroscopic
functional valid beyond local equilibrium   
(where there are no such correlations) is clearly desirable.  
We report here what we believe is the first  exact 
derivation of such a functional for a nonequilibrium model which is
relatively simple but 
exhibits the realistic feature of macroscopically long range correlations.

Before describing our model and results, we review briefly the
corresponding results for equilibrium systems \cite{Lanford,Olla,Ellis,M-L}. 
Let us 
ask for the probability of finding an
isolated macroscopic equilibrium system, having a given number of particles
and energy and contained in a given volume $V$, in a specified 
macro state ${\cal M}$, 
that is, of finding the microscopic configuration $X$ of the system in
a certain region $\Gamma_{\cal M}$ of its phase space.  According to the basic
tenet of equilibrium statistical mechanics, 
embedded in the Boltzmann-Gibbs-Einstein formalism, 
this probability is proportional to $\exp[S({\cal M})/k_B]$, 
where the (Boltzmann) entropy $S({\cal M})$ of the macro
state $\cal M$ is defined by $S({\cal M}) = k_B \log |\Gamma_{\cal M}|$, 
with $|\Gamma_{\cal M}|$ the phase space volume of $\Gamma_{\cal M}$ 
\cite{Lanford,Olla,Ellis,M-L,Cerc}. 

When the system is not isolated but is part of a much larger system---a
situation idealized by considering the system to be 
in contact with an infinite thermal reservoir at
temperature $T$ and chemical potential $\cp$---then the entropy in the
formula above is replaced by an appropriate  free energy.  
Consider in particular a lattice gas in  
a unit cube containing $L^d$  sites 
with spacing $1/L$  (similar formulas will hold for 
continuum systems), and suppose that the macro state of interest  
is specified by a {\it density profile} prescribing the density $\rho(x)$ at
each macroscopic position $x$ in the cube. 
Then the probability of finding the system in this macro state is given for
large $L$ by
 \begin{equation}
  P(\{\rho(x)\}) \sim \exp\big[-L^d{\cal F}_{\rm eq}(\{\rho\})]\;,
  \label{Prob}
 \end{equation}
 with
 \begin{equation}
 {\cal F}_{\rm eq}(\{\rho\})
    =\int\big[f_\cp(\rho(x))-f_\cp(\bar\rho)\big]\,dx\;. \label{Feq}
 \end{equation}
  The integration in (\ref{Feq}) is over the unit cube, 
$f_\cp(r)=a(r)-\cp r$, where $a(r)$ is 
the usual Helmholtz free
energy density for a uniform equilibrium system at density $r$,
and the equilibrium density $\bar\rho=\bar\rho(\cp)$  corresponding to the
chemical potential $\cp$ is obtained by minimizing $f_\cp$:
 \begin{equation}
   \cp=\left.\partial a(r)\over\partial r\right|_{r=\bar\rho(\cp)}.
 \label{cprho}
 \end{equation}
 Note that $-f_\cp(\bar\rho)$ is just the pressure in the grand canonical
ensemble.    We have
suppressed the dependence of $f$ and ${\cal F}_{\rm eq}$ on the constant
temperature $T$, and assume for simplicity that neither $\bar\rho$ nor
$\rho(x)$, for any $x$, lies in a phase transition region at this
temperature. 

The challenge is to extend these results
to nonequilibrium systems, in particular, to systems which are
maintained in a stationary nonequilibrium state (SNS) 
with a steady  flux of particles  by contact
with two boundary reservoirs at different chemical potentials.
We would like to 
generalize the formula (\ref{Feq}) for ${\cal F}_{\rm eq}(\{\rho\})$,
obtaining a large deviation functional 
${\cal F}(\{\rho\})$ such that the
probability of observing a density profile $\rho(x)$ 
is given by a formula analogous to (\ref{Prob}).
The typical profile would then correspond to the $\rho$ which
minimizes $\F$.    A similar analysis and an appropriate 
$\F$ would certainly be useful for the study of
pattern formation in more general SNS.  An interesting example is 
the B\'enard system, in which the
particle  flux is replaced by a heat flux maintained by
reservoirs at different temperatures, with the hotter reservoir below 
the system, and the typical patterns change
abruptly from uniform to rolls to hexagonal cells as the fluid is driven away
from equilibrium \cite{CH}. 
Open systems of this type have been discussed extensively in the literature
from both a microscopic and macroscopic point of view; see \cite{G,E} and
references therein.

Given our current limited understanding of such SNS, however, it is
necessary to start with the simplest systems; our results here are for the
one dimensional symmetric simple exclusion process (SSEP)
\cite{Ligg1,Ligg2,HS}, driven
by boundary reservoirs at distinct chemical potentials $\cp_0$ and $\cp_1$. 
We consider a lattice of $N$ sites, in which each site $i$ is either empty
($\tau_i=0$) or occupied by a single particle ($\tau_i=1$), so that each of
the $2^N$ possible configurations of the system is characterized by $N$
binary variables $\tau_1,...,\tau_N$.  Each particle independently attempts
to jump to its right neighboring site, and to its left neighboring site, in
each case at rate $1$ (so that there is no preferred direction).  It succeeds
if the target site is empty; otherwise nothing happens.  At the boundary
sites, $1$ and $N$, particles are added or removed: a particle is added to
site $1$, when the site is empty, at rate $\alpha$, and removed, when the
site is occupied, at rate $\gamma$; similarly particles are added to site $N$
at rate $\delta$ and removed at rate $\beta$.  This corresponds to the system
being in contact with infinite left and right reservoirs having fugacities
$z_0 = \exp \cp_0 = \alpha / \gamma$, and $z_1 = \exp \cp_1 = \delta/\beta$,
see \cite{ELS1,ELS2}.     We therefore define
 \begin{equation}
 \rho_0={z_0\over1+z_0}={\alpha\over\gamma+\alpha}, \qquad \qquad
  \rho_1={z_1\over1+z_1}={\delta\over\beta+\delta}\;, \label{rho12}
 \end{equation}
 and think of these as the densities of the reservoirs.  They will in fact be
the stationary densities at the left and right ends of the system when
$N\to\infty$ (see (\ref{tau}) below). 

We refer to $z_0$ and $z_1$ as fugacities because 
if we were to place our system in contact with only
the left (right) reservoir by limiting particle input and output to just the
left (right) side, i.e., by setting $\delta=\beta=0$ 
($\alpha=\gamma=0$), then its
stationary state would be one of equilibrium with fugacity $z_0$ ($z_1$),
with no net flux of particles.  Of
course if $z_0=z_1=z$ then the system would be in an equilibrium state 
whether in contact with one or both reservoirs, i.e., in 
a product measure with uniform density
$\rho=z/(1+z)$;  this value follows from (\ref{cprho}), since for this system 
 \begin{equation}
  a(r)=r\log r+(1-r)\log(1-r).\label{arho}
 \end{equation}

For the model considered here the typical
profile $\bar\rho(x,\tilde t)$, on the macroscopic spatial-temporal scale with
variables $(x,\tilde t)$ defined by
$i\to xN$ and $t\to \tilde tN^2$, is for $N\to\infty$ governed by the
diffusion equation \cite{ELS1,ELS2,KL}
 \begin{equation}
   {\partial\bar\rho(x,\tilde t)\over\partial \tilde t}=
     {\partial^2\bar\rho(x,\tilde t)\over\partial x^2}, 
   \qquad \rho(0,\tilde t)=\rho_0, \quad \rho(1,\tilde t)=\rho_1,
 \end{equation}
 which gives for the stationary state
$\bar\rho(x)=\rho_0+(\rho_1-\rho_0)x$.  
Spohn \cite{HS1} (see also \cite{DFIP}) 
computed explicitly, for $\alpha+\gamma=\beta+\delta=1$, both
the expected density profile and the pair correlation in the
stationary state, and showed that the results agree, in the above scaling
limit, with those obtained from fluctuating hydrodynamics.

While higher order correlations can also   be obtained in principle, 
the difficulty of the computation increases rapidly with the
desired order.  The complete measure on the microscopic configurations in
the steady state may, however, be computed
through the so-called matrix method \cite{DEHP,ER,RS,Sas,BECE}.  
From this measure we
would like to determine the probability of seeing an arbitrary macroscopic
density profile $\rho(x), x \in [0,1]$; by definition, this is the sum of
the probabilities of all microscopic configurations which are consistent
with $\rho(x)$,  e.g.,
all configurations $\tau$ such that for any $x_0,x_1$ with $0\le x_0<x_1\le1$,
$\left|{1
\over N} \sum_{i= x_0 N}^{ x_1 N}\tau_i-\int_{x_0}^{x_1}
\rho(x)\,dx\right|<\delta_N$, 
for some appropriate choice of $\delta_N$ with
$\delta_N\to0$ as $N\to\infty$, 
e.g., $\delta_N=N^{-\kappa}$, $0<\kappa<1$.

The deviations of macroscopic systems from typical behavior, e.g., from the
profile $\bar\rho(x,\tilde t)$ governed by the diffusion equation, is a
central issue of statistical mechanics \cite{Lanford,Olla,Ellis,M-L,G,E}. 
The time dependent  problem was first
studied   by Onsager and Machlup \cite{OM}, who
considered space-time fluctuations for Hamiltonian systems 
in equilibrium (for which
$\bar\rho$ is constant).  The problem of observing a profile
$\rho(x,\tilde t)$ not necessarily close to $\bar\rho(x,\tilde t)$, the
so-called problem of large deviations, was later studied rigorously for the
SSEP on a torus by Kipnis, Olla, and Varadhan \cite{KOV}, who found an
exact expression for the probability of observing such a profile  (see
also \cite{HS,KL}).  An
important ingredient in their analysis was the fact that the dynamics
satisfied detailed balance with respect to any product measure
with constant density, and that such a product measure
is in fact the appropriate microscopic measure at each
macroscopic position.  This corresponds for stochastic systems to
reversibility of the microscopic dynamics, which also 
plays an important role in
the Onsager-Machlup theory.
There have been various attempts to extend these
results to open systems in which the dynamics no longer satisfy detailed
balance with respect to the stationary measure
\cite{E,ELS3,BDGJL,BDGJL2}.  Bertini et al.~\cite{BDGJL} have
indeed succeeded in doing this for the so-called zero-range process with open
boundaries, for which the microscopic stationary state is also a product
measure.  The problem is more difficult for our system, where it is
known \cite{HS1} 
that the stationary microscopic state contains long range
correlations---correlations expected to be generic for SNS 
\cite{Sch,GLMS,Grin,ZS} and
even measured experimentally in some cases \cite{Exp}.  We shall see that
this is reflected in a nonlocal structure for $\F(\{\rho\})$.  A preliminary
version of our results was given in \cite{DLS}, and more recently Bertini et
al. \cite{BDGJL2} succeeded in rederiving this result, and obtaining a large
deviation functional for space-time profiles, using a semi-macroscopic 
method (which could be further checked,
at least for small fluctuations, by comparing its predictions with
the direct calculation of the time dependent correlations given in 
\cite{schutz}).

\section{Summary of results\label{summary}}

In the present paper we give the exact asymptotic formula for the probability
$P_N(\{\rho(x\})$ of seeing a density profile $\rho(x)$, $0 \leq \rho(x) \leq
1$, in the one dimensional open system with SSEP internal dynamics.  (The
exact sense in which we establish this formula is sketched later in this
section.)  Let $\rho_0$ and
$\rho_1$ be defined as in (\ref{rho12}); we will assume for definiteness that
$\rho_0>\rho_1$   
(results for the case $\rho_0<\rho_1$ then 
follow from the right-left symmetry or 
the particle-hole symmetry, and for the case $\rho_0= \rho_1$
by taking the limit $\rho_0 \to \rho_1$).

Our main result is then:
 \begin{equation}
   \lim_{N\to \infty}{\log P_N(\{\rho(x\}) \over N}
  \equiv -{\cal F}(\{\rho\})\;,
 \end{equation}
 where 
 \begin{eqnarray}
 {\cal F}(\{\rho\})
 &=& \int_0^1 dx \left\{ \rho(x) \log \left({ \rho(x) \over F(x)}\right) 
   + (1- \rho(x)) \log \left( 1 - \rho(x) \over 1 -F(x) \right)\right.
     \nonumber\\
 && \hskip100pt 
  + \left.\log \left({ F'(x) \over \rho_1 - \rho_0} \right) \right\}.
\label{simple_expression}
\end{eqnarray}
 The auxiliary function $F(x)$ in (\ref{simple_expression}) is given as a
function of the density profile $\rho(x)$ by the monotone solution of the
nonlinear differential equation
 \begin{equation}
\rho(x) =
 F(x) + {F(x)(1-F(x)) F''(x) \over F'(x)^2}\; ,
\label{rho(t)}
 \end{equation}
with the boundary conditions
 \begin{equation}
 F(0)= \rho_0 \;,  \ \ \ \ \; \ \ \ \  F(1)= \rho_1 \;.
\label{condition5} 
 \end{equation}
We will show in Section~\ref{continuum} that such a solution exists and is
(at least when $\rho_0<1$ and $\rho_1>0$) unique.

We now summarize some  consequences of 
(\ref{simple_expression}--\ref{condition5}).

 (a) Let us denote the right hand side of (\ref{simple_expression}),
considered as a functional of two 
independent functions $\rho(x)$ and $F(x)$, by $\G$:
 \begin{eqnarray}
 \G(\{\rho\},\{F\})) &=&
   \int_0^1 dx \left\{ \rho(x) \log \left({ \rho(x) \over F(x)}\right) 
   + (1- \rho(x)) \log \left( 1 - \rho(x) \over 1 -F(x) \right)\right.
     \nonumber\\ && \hskip100pt 
  + \left.\log \left({F'(x) \over \rho_1 - \rho_0} \right) \right\}.
\label{defG}
\end{eqnarray}
 If one looks for a monotone function $F$, satisfying the constraint
(\ref{condition5}), for which ${\cal G}(\{\rho\},\{F\})$ is an extremum, one
obtains (\ref{rho(t)}) as the corresponding  Euler-Lagrange equation:
 \begin{equation}
{\delta {\cal G}(\{\rho\},\{F\}) \over \delta F(x) } = 0.
  \label{Euler-Lagrange}
 \end{equation}
  We will show in Section~\ref{continuum} 
that the unique monotone solution of 
(\ref{rho(t)}) (or equivalently of (\ref{Euler-Lagrange}))
is in fact a maximizer of (\ref{defG}):
 \begin{equation}
{\cal F(\{\rho\})}
  = \sup_F{\cal G}(\{\rho\},\{F\})\;.
 \label{GF}
 \end{equation}
 We will also show that $\F$ is a convex function of $\{\rho\}$. (More
precisely, our derivations will be for the case $1>\rho_0>\rho_1>0$, but we
expect the conclusions to hold in general, that is, for
$1\ge\rho_0>\rho_1\ge0$ and by symmetry for $1\ge\rho_1>\rho_0\ge0$.) 

(b) From (\ref{Euler-Lagrange}) one sees at once that 
 for $F$ the solution of (\ref{rho(t)}), (\ref{condition5}),
 \begin{equation}
{\delta {\cal F}(\{\rho\}) \over \delta \rho(x)}
  ={\delta {\cal G}(\{\rho\},\{F\}) \over \delta \rho(x)}
  = \log\left[{\rho(x) \over F(x)}
  \cdot {1-F(x) \over 1-\rho(x)}\right],\label{vd}
 \end{equation}
 and this together with (\ref{rho(t)}) implies 
that the minimum of $\F(\{\rho\})$ occurs for
$\rho(x)=F(x)=\bar\rho(x)$, where 
 \begin{equation}
\bar\rho(x)=\rho_0(1-x)+\rho_1x. \label{rhobar}
 \end{equation}
  Moreover, from (\ref{simple_expression}) and (\ref{rho(t)}) one has 
$\F(\{\bar\rho\})=0$, confirming that the most likely profile
$\bar\rho(x)$ is obtained with probability one in the limit $N\to\infty$. 
Any other profile will have ${\cal F}(\{\rho\}) > 0$ and thus, for large $N$,
exponentially small probability.  
The profile may be discontinuous, or may fail to satisfy
the boundary conditions $\rho(0)=\rho_0$ or $\rho(1)=\rho_1$,
and still satisfy ${\cal F}(\{\rho\}) < \infty$.
When $\rho_0 = 0$ or $\rho_1 = 1$ there are
some profiles for which ${\cal F} = +\infty$; their probability is
super-exponentially small in $N$.  For examples, see (d) below and
Section~\ref{examples}.

 (c) It is natural to contrast the SNS under consideration here with a {\it
local equilibrium} Gibbs measure for the same system---a 
lattice gas with only hard core exclusion---with no reservoirs at the
boundaries but with a spatially varying chemical
potential $\cp(x)$ \cite{KOV,HS} which is adjusted to maintain the same
optimal profile $\bar\rho(x)$.  For this system the large deviation
functional (free energy) is obtained directly from (\ref{Feq}), with
$f_\cp(r)=r\log r+(1-r)\log(1-r)-\cp r$:
 \begin{equation}
 {\cal F}_{\rm eq}(\{\rho\})
 =  \int\bigg\{ \rho(x) \log {\rho(x) \over \bar \rho(x)}
     +[1- \rho(x)]\log{(1 - \rho(x))\over(1 - \bar \rho(x))}\bigg\}\,dx.
   \label{equil}
 \end{equation}
 In general the two  expressions (\ref{simple_expression})
and (\ref{equil}) are different, and from (\ref{GF}),
 \begin{equation}
 {\cal F}(\{\rho\})
  = \sup_F{\cal G}(\{\rho\},\{F\}) 
  \ge  {\cal G}(\{\rho\},\{\bar\rho\})
  ={\cal F}_{\rm eq}(\{\rho\})\;.\label{FFeq}
 \end{equation}
 $\F(\{\rho\})$ and $\Feq(\{\rho\})$ agree only for
$\rho(x)=\bar\rho(x)$ or in the limiting case $\rho_0=\rho_1$, 
in which the system is in equilibrium with
$\bar\rho(x)=\rho_0$.  (Equations (\ref{simple_expression}--\ref{condition5}) 
have a well-defined limit for $\rho_1\nearrow\rho_0$, with 
$F(x)=\rho_0 + (\rho_1 - \rho_0) x + O( (\rho_1 - \rho_0)^2) $.)  
Otherwise
$\F(\{\rho\})$ lies above $\Feq(\{\rho\})$ and thus gives reduced probability
for fluctuations away from the typical profile.

Note that the integrand in (\ref{equil}) (or (\ref{Feq})) is local: changing
$\rho(x)$ in some interval $[a,b]$ only changes the value of this integrand
inside that interval.  This is not true for the integrand of
(\ref{simple_expression}), because $F$ is determined by the differential
equation (\ref{rho(t)}) and so $F(x)$ will generally depend on the value of
$\rho(y)$ everywhere in $[0,1]$.

 (d) For a constant profile $\rho(x)=r$, 
the solution $F$ of (\ref{rho(t)}) and (\ref{condition5}) 
satisfies $F'= A F^r(1-F)^{1-r}$, where  $A$ is
fixed by (\ref{condition5}), and
 \begin{equation}
{\cal F}(\{\rho\})  = \log \left[  \int_{\rho_0}^{\rho_1} \left( r 
\over z \right)^r \  \left( 1-r \over 1 - z \right)^{1-r} \ {dz \over 
\rho_1 - \rho_0} \right]\;.
\label{constantrho}
 \end{equation}
We see that ${\cal F}(\{\rho\})= \infty$
if $r=0$ and $\rho_0=1$, or $r=1$ and $\rho_1=0$. 
By contrast $\Feq(\{\rho\})$ as given in (\ref{equil}) would be
 \begin{equation}
\Feq(\{\rho\})  = \int_{\rho_0}^{\rho_1} \log\left[\left( r 
\over z \right)^r \  \left( 1-r \over 1 - z \right)^{1-r}\right] \ {dz \over 
\rho_1 - \rho_0} \;,
\label{eqconstantrho}
 \end{equation}
 which is finite except in the degenerate cases $\rho_0=\rho_1=1$ and
$\rho_0=\rho_1=0$,  
 in which case the
measure is concentrated on a single configuration and $\Feq=\infty$ unless
$\rho=\rho_0$. 

 (e) If we minimize ${\cal F}$ subject to the constraint of a fixed mean
density $\int_0^1 \rho(x)dx$, the right hand side of (\ref{vd}) becomes an
arbitrary constant, and together with (\ref{rho(t)}) one obtains that the
most likely profile is exponential: $\rho(x)=A_1\exp(\theta x)+ A_2$ (with
$F(x)=1-A_2+A_2(1-A_2)\exp(-\theta x)/A_1$), the
constants being determined by the value of the mean density and the boundary
conditions (\ref{condition5}).  (This exponential form, which is the
stationary solution of a diffusion equation with drift, was first suggested
to us by Errico Presutti). 

Similarly, if we impose a fixed mean density in $k$ nonoverlapping intervals,
$\int_{a_i}^{b_i} \rho(x)dx = c_i$ for $i=1,\ldots,k$, with no other
constraints, then one can show using (\ref{rho(t)}) and (\ref{vd}) that the
optimal profile has an exponential form inside these intervals and is linear
outside; it will in general not be continuous at the end points of the
intervals. 

(f) Using the fact that the exponential is the optimal profile for the case
of a fixed mean density in the entire interval, we may compute the
distribution of $M$, the total number of particles in the system, in the
steady state for large $N$.  We find that the fluctuations of $M$ predicted
by (\ref{simple_expression}) are reduced in comparison to those in a system
in local equilibrium (\ref{equil}) with the same $\bar\rho$:
 \begin{eqnarray}
  \lim_{N\to\infty}N^{-1}
    \bigl[\langle M^2\rangle_{\rm SNS}-\langle 
M\rangle^2\bigr]\nonumber\\
 &&\hskip-115pt =\;
   \lim_{N\to\infty}N^{-1}
    \bigl[\langle M^2\rangle_{\rm eq}-\langle M\rangle^2\bigr]
      - {(\rho_1 - \rho_0)^2 \over 12} .
\label{variance}
\end{eqnarray}
   (Since $\bar\rho$ is given by (\ref{rhobar}) for both systems, 
   $\langle M\rangle_{\rm SNS}=\langle M\rangle_{\rm eq}
      =\langle M\rangle=(\rho_0+\rho_1)/2$.)
 This reduction of fluctuations was already visible in (\ref{FFeq}).
 We may also obtain (\ref{variance}) by expanding $\rho(x)$ about
$\bar\rho(x)$ in (\ref{simple_expression}).  The result agrees with that
obtained in \cite{HS1} directly from the microscopic model and from
fluctuating hydrodynamics \cite{DKS}.

The structure of the rest of the paper is as follows: In Sections
\ref{genfct} and \ref{gfld} and Appendices \ref{algebra} and
\ref{asymptotics} we derive our main result
(\ref{simple_expression}--\ref{condition5}) from the knowledge of the
weights of the microscopic configurations as given by their matrix product
expressions.  Our approach consists in considering the system of $N$ sites as
decomposed into $n$ boxes of $N_1,N_2,\ldots,N_n$ sites.  We first calculate
the generating function of the probability
$P_{N_1,\ldots,N_n}(M_1,\ldots,M_n)$, that $M_1$ particles are located on the
first $N_1$ sites, $M_2$ particles on the next $N_2$ sites of the lattice, etc. 
We then use this generating function to obtain
$P_{N_1,\ldots,N_n}(M_1,\ldots,M_n)$.  The expression we obtain appears in a
parametric form because we extract this probability from the generating
function through a Legendre transformation.  Then we take the limit of an
infinite system which corresponds implicitly to the usual hydrodynamic
scaling limit, as explained in great detail in \cite{HS,KL}, 
by first letting $N \to \infty$, keeping $N_i/N = y_i$ fixed, and then 
letting $y_i \to 0$
to obtain (\ref{simple_expression}--\ref{condition5}). 

In Section \ref{continuum} we prove that for any profile $\rho(x)$ there
exists a unique monotonic function $F(x)$ which satisfies (\ref{rho(t)})
and (\ref{condition5}).  We also establish there, for $1>\rho_0>\rho_1>0$,
that $\F(\{\rho\})=\sup_F\G(\{\rho\},\{F\})$ and that $\F$ is a convex
functional of $\rho(x)$, as discussed in (a) above.  In the course of this
discussion we describe the behavior of the large deviation functional for
piecewise constant density profiles.  In Section \ref{optprof} we calculate
optimal profiles under various constraints, as discussed in (e) above.  In
Section \ref{fluctuations} we calculate the correlations of the
fluctuations of the density profile around the most probable one and we
show that a direct calculation of these correlations agrees with what can
be calculated from (\ref{simple_expression}--\ref{condition5}).  Lastly in
Section \ref{examples} we exhibit a few examples of density profiles for
which one can calculate explicitly the function $F$ and the value of
${\cal F}(\{\rho\})$.

\section{Exact generating function\label{genfct}}

For the SSEP with open boundaries as described in the Introduction, the
probability of a configuration $\tau = \{\tau_1,\dots,\tau_N\}$ in the
 (unique) steady state of our model is given by \cite{DEHP,ER,RS,Sas,BECE},
\cite{Ligg2},
 \begin{equation}
   P_N(\tau) = {\omega_N(\tau)
    \over  \langle W|(D+E)^N|V\rangle}\;, \label{prob}
 \end{equation}
where the weights $\omega_N(\tau)$ are given by
 \begin{equation}
   \omega_N(\tau) = \langle W|\Pi_{i=1}^N (\tau_i D + (1 - 
\tau_i)E)|V\rangle 
    \label{omega}
 \end{equation}
  and the matrices $D$ and $E$ and the vectors $|V\rangle$ and $\langle 
W|$ satisfy
 \begin{eqnarray}
     DE - ED &=& D + E  \label{DE}\;,\label{alg1}\\
  (\beta D - \delta E)|V\rangle &=& |V\rangle\;,\label{alg2}\\
   \langle W|(\alpha E - \gamma D) &=& \langle W|\;\label{alg3}. 
 \end{eqnarray}

To obtain the probability of a specified density profile we first calculate
the sum $\Omega_{N_1,\dots,N_n}(M_1,M_2,\dots,M_n)$ of the weights $\omega_N$
of all the configurations with $M_1$ particles located on the first $N_1$
sites, $M_2$ particles on the next $N_2$ sites, etc.  The key to obtaining
(\ref{simple_expression}) is that the following generating function
can be computed exactly:
\begin{eqnarray}
 Z(\fp_1,\ldots,\fp_n; \mu_1,\ldots, \mu_n) &&\nonumber \\
 &&\hskip-70pt\equiv
  \sum {\mu_1 ^{N_1}\over N_1!}\cdots{\mu_n^{N_n}\over N_n!}
 \fp_1 ^{M_1} \cdots \fp_n ^{M_n} 
{\Omega_{N_1,\dots,N_n}(M_1,\ldots,M_n) \over \langle W | V \rangle}
   \nonumber\\
 &&\hskip-70pt =
 { \langle W | e^{\mu_1 \fp_1 D + \mu_1 E}  \cdots e^{\mu_n 
\fp_n D + \mu_n E} | V \rangle  \over
\langle W |  V \rangle }\;,
\label{Zdef}
 \end{eqnarray}
where  the sum is over all $N_i,M_i$ with $0\le M_i\le N_i$.
As shown in Appendix~\ref{algebra}, $Z$ can be computed explicitly:
 \begin{eqnarray}
 Z = \left(\rho_0 - \rho_1 \over g \right)^{a+b} \exp \left[ a 
\sum_{i=1}^n \mu_i(1-\fp_i) \right] \;,
\label{Zresult}
 \end{eqnarray}
where $\rho_0$ and $\rho_1$ are given by (\ref{rho12}),
 \begin{equation}
 a={1\over\gamma+\alpha}, \qquad b={1\over\beta+\delta}\;, \label{abdef}
 \end{equation}
 and  
 \begin{eqnarray}
 g &=& -\rho_1 + \rho_0 e^{\sum_{i=1}^n \mu_i(1-\fp_i)} 
\nonumber\\
 &&\hskip45pt  +\sum_{i=1}^n {1 \over \fp_i - 
1}(e^{\mu_i(1-\fp_i)} - 1)
e^{\sum_{j>i} \mu_j(1-\fp_j)}.
\label{gdef}
 \end{eqnarray}
 Expressions (\ref{Zresult}) and (\ref{gdef}) are the basis of all
the calculations leading to the large deviation functions.  As a first
step, note that 
 \begin{equation}
  Z(1; \mu) 
   = { \langle W | e^{\mu (D+E)}  | V \rangle  \over
    \langle W |  V \rangle } 
  = \left( \rho_0 - \rho_1 \over \rho_0 - \rho_1 - \mu\right)^{a+b},
 \label{Z111}
 \end{equation}
 so that  one obtains the normalization factor in  (\ref{prob}): 
 \begin{equation}
  \Omega_0\equiv { \langle W | (D+E)^N | V \rangle  \over
   \langle W |  V \rangle } 
  =  {\Gamma(a+b+N) \over \Gamma(a+b) (\rho_0 - \rho_1)^N} \;.
\label{Z0}
 \end{equation}

\section{From the generating function to the large deviation function
  \label{gfld}}

It is clear from (\ref{Zresult})
that $Z(\fp_1, \cdots \fp_n; \mu_1,\cdots \mu_n)$ is singular on
the hypersurface
 \begin{equation}
g(\mu_1,\cdots, \mu_n; \fp_1, \cdots, \fp_n)=0,
\label{feq}
 \end{equation}
 where $g$ is given by (\ref{gdef}).
 Let us consider a very large system of $N$ sites divided 
into $n$ boxes of $N_1,\ldots,N_n$ sites, 
with $ P_{N_1,... N_n} (M_1, ... M_n)$  the probability of finding
$M_1$ particles in the first box, $M_2$ particles in the second box, etc., 
and let $N_i$ and $M_i$ be proportional to $N$:
$N_i= N y_i$, $M_i = r_iN_i=r_iy_iN$, $i=1,\ldots,n$.
As explained in  Appendix 2, one can show from the definition of $Z$  
that for large $N$ and fixed $y_i$, $r_i$,  
 \begin{eqnarray}
{\log   P_{N_1,... N_n} (M_1, ... M_n) \over N} && \nonumber\\
 &&\hskip-60pt \simeq\;
   \log(\rho_0 - \rho_1) - \sum_{j=1}^n y_j   \left( \log {\mu_j \over y_j}
     +  r_j \log \fp_j  \right)\;,
\label{Pro}
\end{eqnarray}
  where the box sizes $y_j$ and their particle densities  $r_j$
are related to the  parameters $\mu_1,.. \mu_n$, $\fp_1,...\fp_n$
by
 \begin{eqnarray}
y_j &=&  {   \ {\partial g \over \partial \log \mu_j} 
\over \sum_{i=1}^n  {\partial g \over \partial \log \mu_i} } \;,
 \label{xj}\\
  r_j &=&  {   \ {\partial g \over \partial \log \
\fp_j} 
\over  {\partial g \over \partial \log \mu_j} } \;,
\label{rj}
\end{eqnarray}
 with all derivatives calculated on the manifold $g=0$.

Equation (\ref{Pro}) gives the large deviation function in a parametric form;
the $2 n+1 $ equations (\ref{feq}), (\ref{xj}), and (\ref{rj}) determine the
$2 n$ parameters $\mu_1,...,\mu_n$, $\fp_1,...,\fp_n$ in terms of
$y_1,\ldots,y_n$ and $r_1,\ldots,r_n$ (since $y_1+\cdots+ y_n=1$, the $n$
equations (\ref{xj}) give only $n-1$ independent conditions, so that the
system is not overdetermined).  Note from (\ref{Zresult}) and (\ref{gdef}) 
that the parameters $a$ and $b$
defined by (\ref{abdef})  do not appear in the expression for the
critical manifold and therefore drop out in the large $N$ limit, i.e., the
large deviation functional,  like the typical profile, depends only on
$z_0$ and $z_1$ or $\rho_0$ and $\rho_1$.

\subsection{Case of a single box}

One can apply the above  results in the case $n=1$ of a single box.
With $\fp_1 \equiv \fp$ and $\mu_1 \equiv \mu$,
equation (\ref{feq}) for the critical manifold becomes
 \begin{equation}
 g(\mu;\fp) 
\equiv -\rho_1 +\rho_0 e^{\mu(1-\fp)} + {1 \over \fp -1} 
  ( e^{\mu(1 - \fp)} -1 )=0, 
 \end{equation}
 or more conveniently $\tilde g(\mu;\fp)=0$, where 
 \begin{equation}
\tilde g(\mu;\fp)
 = \mu(1-\fp) -  \log \left({ 1 - \rho_1 + \fp \rho_1 \over 1 -
   \rho_0 + \fp \rho_0 }\right). 
 \end{equation}
 The function $\tilde g$ may be used in place of $g$ in 
(\ref{xj}--\ref{rj}),   because the derivatives there are evaluated on the
manifold $g=0$, so that if the average
density $\rho=r_1$ in the box is given by
 \begin{equation}
\rho
  =  { - {\fp \over 1 - \fp} 
   \log \left({ 1 - \rho_1 + \fp \rho_1 \over 1 - \rho_0 + \fp
         \rho_0 }\right) 
      + { \rho_0 \fp  \over 1 - \rho_0 + \fp \rho_0} 
      - {\rho_1 \fp \over 1 - \rho_1 + \fp \rho_1} \over
      \log \left({ 1 - \rho_1 + \fp \rho_1 
            \over 1 - \rho_0 + \fp \rho_0 }\right)} \;,
\label{singlerho}
 \end{equation}
 then  for large $N$ and $M\simeq \rho N$, 
 \begin{equation}
 {1\over N}\,{\log P_N(M)} 
  \simeq - \rho \log \fp \  - 
 \  \log \left[ {\log \left({ 1 - \rho_1 + \fp \rho_1 
   \over 1 - \rho_0 + \fp \rho_0 }\right) 
   \over (\fp-1) (\rho_1 - \rho_0) } \right].
\label{singlepro}
 \end{equation}

{ \bf Remark}: For small fluctuations of the form
 \begin{equation}
\rho - {\rho_0 + \rho_1 \over 2 } = \delta  \rho \ll 1 \;,
 \end{equation}
 (\ref{singlepro}) gives, again with $M\simeq \rho N$, 
 \begin{equation}
  {1\over N}\,{\log P_N(M)} 
   \simeq  { -6 \ (\delta \rho)^2 \over  
   6 \rho_0 + 6 \rho_1 - 5 \rho_0^2  - 2 \rho_0 \rho_1 - 5 \rho_1^2}  
   + O\left((\delta \rho)^3\right)\;.
 \label{fluct}
 \end{equation}
 For  a Bernoulli distribution with average profile 
(\ref{rhobar})  (the most likely profile, as we will see in 
 Section~\ref{optprof}),
one would get
 \begin{equation}
   {1\over N}\,{\log P_N(M)} 
   \simeq  { -3 \  (\delta \rho)^2 \over  3 \rho_0 
  + 3 \rho_1 - 2 \rho_0^2  - 2 \rho_0 \rho_1 - 2 \rho_1^2} 
  +  O\left((\delta \rho)^3\right)  \;;
 \end{equation}
  from this we can compute the fluctuations of $M$ and obtain 
(\ref{variance}).

\subsection{A finite number $n$ of large boxes} 

Let us calculate ${\partial g\over \partial \log \mu_i}$ 
on the manifold $g=0$.  From (\ref{gdef}), we
have 
 \begin{eqnarray}
 {\partial g \over \partial \log \mu_i} = \rho_0 \mu_i
(1- \fp_i) e^{\sum_{j=1}^n \mu_j (1 - \fp_j)} - \mu_i e^{\sum_{j=i}^n
\mu_j (1 - \fp_j)}  \nonumber \\ 
   + \mu_i (1 - \fp_i)  \sum_{j<i}^n {1 \over
 \fp_j - 1} (e^{\mu_j (1 - \fp_j)} -1) e^{ \sum_{k >j} \mu_{k }
  (1 - \fp_{k })} .
 \end{eqnarray}
which becomes, using   (\ref{feq}) and (\ref{gdef}),
 \begin{eqnarray}
{\partial g \over \partial \log \mu_i}
&=& - \mu_i  e^{\sum_{j>i} \mu_j (1 - \fp_j)}
  \nonumber \\
 &&\hskip-55pt + \mu_i (1 - \fp_i)  \sum_{j>i} {1 \over
1-  \fp_j } (e^{\mu_j (1 - \fp_j)} -1) e^{ \sum_{k >j} \mu_{k } (1 - \fp_{k })} 
  +  \rho_1 \mu_i (1 - \fp_i) \;.\ \ \ \ \ \ \   
\label{dmexp}
 \end{eqnarray}
 One can also  show that
 \begin{eqnarray}
{\partial g \over \partial \log \fp_i}
= - {\fp_i \over 1 - \fp_i} {\partial g \over \partial \log \mu_i}
- {\fp_i \over (\fp_i-1)^2}
 (e^{\mu_i (1 - \fp_i)} -1) e^{ \sum_{j>i} \mu_{j} (1 - \fp_{j})}\;.
\label{dlexp}
\end{eqnarray}

This parametric form can be simplified by  replacing the role 
of the two sequences of parameters $\mu_i$ and $\fp_i$ by a single sequence  of parameters $G_i$.
Let us define the constant $C$ by
 \begin{equation}
C= \sum_{i=1}^n {\partial g \over \partial \log \mu_i} \;,
\label{Cdef}
 \end{equation}
and the sequence $G_i$ by
 \begin{equation}
G_i =  {1 \over C}   e^{\sum_{j=i}^n \mu_j (1 - \fp_j)} \ \ \ \ , 
    \ \ \ \  G_{n+1} = {1 \over C} \;.
\label{Gdef}
 \end{equation}
 From the very definition of the $G_i$'s it is clear that 
 \begin{equation}
\mu_i= {\log(G_i/G_{i+1}) \over 1 - \fp_i} \;,
\label{muisol}
 \end{equation}
and from (\ref{xj}), (\ref{dmexp}), (\ref{Cdef}), and (\ref{Gdef}), 
we see that
 \begin{equation}
{G_{i+1} \over 1 -\fp_i} 
= - {  y_i \over \log(G_i/ G_{i+1})} + \sum_{j>i} {G_j - G_{j+1} 
   \over 1 - \fp_j}  + { \rho_1 \over C}\;,
\label{lambdaisol}
 \end{equation}
so that both the $\mu_i$ and the $\fp_i$ are determined in terms of the 
sequence
$G_i$,  the box sizes $y_i$ and the constant $C$.
The condition (\ref{feq}) that $g=0$ becomes
 \begin{equation}
{\rho_1 \over C} + \sum_{i=1}^n {G_i - G_{i+1} \over 1 - \fp_i} = \rho_0 G_1\;,
\label{condition}
 \end{equation}
which gives the extra equation needed to determine the constant $C$ 
(as well as $G_{n+1}$). Therefore we are left with $n$ free parameters, the $G_i$'s for $1 \leq i \leq n$.

A more convenient way of writing the $\fp_i$'s 
(i.e. (\ref{lambdaisol}))
in terms of the $G_i$'s is
 \begin{equation}
{1 \over \fp_i -1} = 
{ 1 \over G_{i+1}} {y_i \over \log(G_i/G_{i+1})} - \rho_0 + \sum_{j=1}^i 
\left( {1 \over G_j} - {1 \over G_{j+1}} \right) {y_j \over \log(G_j/G_{j+1})}\;, 
\label{lambdaisol1}
 \end{equation}
and the condition (\ref{condition}) becomes
 \begin{equation}
\rho_0 - \rho_1 = \sum_{j=1}^n
\left( {1 \over G_j} - {1 \over G_{j+1}} \right) {y_j \over \log(G_j/G_{j+1})}\;,
\label{condition1}
 \end{equation}
which can be thought as an equation which determines $G_{n+1}$ in terms of
the $G_i$'s.

Once the $\fp_i$ are known through  (\ref{lambdaisol1}) and 
(\ref{condition1}),
one gets from (\ref{rj}), (\ref{dmexp}), and (\ref{dlexp}) 
  expressions for the $r_i$'s and the large deviation function:
 \begin{equation}
r_i = - {\fp_i \over 1 - \fp_i} - { \fp_i \over (1- \fp_i)^2} \ { G_i - G_{i+1} \over y_i} \;,
\label{ri1}
 \end{equation}
and
 \begin{eqnarray}
\label{Pro1}
{\log \left[ P_{N_1,... N_n} (M_1, ... M_n)\right] \over N} \\
 &&\hskip-100pt \simeq\;
  - \sum_{i=1}^n \left\{ y_i \log\left[ {\log \left( G_{i} \over G_{i+1} \right) 
\over y_i (\rho_0 - \rho_1)} \right] 
 - y_i \log(1-\fp_i) 
+ y_i  r_i \log(\fp_i) 
\right\}\;.
\nonumber 
\end{eqnarray}
 Equations (\ref{lambdaisol1}--\ref{Pro1}) determine the large deviation
function for an any specified number $n$ of large boxes, i.e.\ for fixed
$y_i$, $r_i$, and $N \to \infty$. 

\subsection{An  infinite  number of large boxes and the continuous limit}

Letting $n$ become large while keeping  each box a small fraction of the
total system, i.e., all the $y_i$'s are small or, more formally letting $N
\to \infty$ followed by $n \to \infty$ and $y_i \to 0$, 
one can introduce a continuous variable
$x$,  $0 \leq x \leq 1$ and let 
 \begin{equation}
x_i= y_1 + y_2 + ...+ y_i\;.
\label{tdef}
 \end{equation}
All the discrete sequences can now be thought of as functions of $x$ with 
 \begin{equation}
G_i \equiv G(x_i) \ \ \ ; \ \ \ \fp_i \equiv \fp(x_i) \ \ \ ; \ \ \
r_i \equiv \rho(x_i), \ \ \ i=1,...,n \;. \label{continuouslimit}
 \end{equation}
Then using extrapolations to make $G, \fp, \rho$ smooth functions of
$x$ so that 
 \begin{equation}
G_{i+1} - G_i \simeq y_i G'(x), 
 \end{equation}
 one finds that (\ref{lambdaisol1}) and (\ref{condition1}) become
 \begin{equation}
{1 \over \fp(x) -1} = {-1 \over G'(x)} - \rho_0 - \int_0^x {du \over G(u)}\;,
\label{lambdasol2}
 \end{equation}
 \begin{equation}
 \rho_0 + \int_0^1 {du \over G(u)} = \rho_1 \;.
\label{condition2}
 \end{equation}
At this stage it is convenient to replace the function
$G(x)$ by another function $F(x)$ defined by
 \begin{equation}
F(x) =  \rho_0 + \int_0^x {du \over G(u)} \;.
\label{Fdef}
 \end{equation}
The expression (\ref{lambdasol2}) of $\fp(x)$ becomes then  
 \begin{equation}
{1 \over \fp(x) -1} =  { F'^2(x) \over F''(x)} - F(x) \;.
\label{lambdasol3}
 \end{equation}

 Using the above relations
we may rewrite (\ref{ri1}) and (\ref{Pro1}) in terms of $F$,
obtaining
(\ref{simple_expression}) and (\ref{rho(t)}); 
the boundary conditions (\ref{condition5}) come from
(\ref{condition2}).  The monotonicity of $F$ follows from the
uniqueness of the sign of
$G$ (see  (\ref{Gdef});  note from (\ref{Fdef}) that $G$ is 
negative in the case $\rho_0 > \rho_1$ that we consider).

\section{The large deviation functional in the continuum limit
 \label{continuum}}

We derive here some properties of equations
(\ref{simple_expression}--\ref{condition5}).  We discuss only the case
$1>\rho_0>\rho_1>0$ and concentrate on results for piecewise constant density
profiles, in some instances giving only a sketch of the extension to more
general $\rho(x)$. The case in which either $\rho_0=1$ or $\rho_1=0$ seems
technically more difficult; for this case we can show the existence of a
solution, but omit the proof here. 

\subsection{Uniqueness}

 In this section we show that any monotone solution $F(x)$ of (\ref{rho(t)}) and
(\ref{condition5}) is unique.  If $F$ and $\hat F$ are distinct solutions
then necessarily $F'(0)\ne\hat F'(0)$, since the standard initial value
problem for (\ref{rho(t)}), with prescribed values of $F(0)$ and $F'(0)$, 
 has a unique solution.  Suppose that $\hat F'(0)>F'(0)$;
we will show that then $\hat F(x)>F(x)$ for $0<x\le1$, contradicting
$F(1)=\hat F(1)=\rho_1$.  For otherwise let $y$
to be the smallest positive number for which $F(y)=\hat F(y)$,
so that
 \begin{equation}
F(x)<\hat F(x)\;, \quad 0<x<y\le1.\label{ineq}
 \end{equation}
 Let $g(x)=F(x)(1-F(x))/F'(x)$ and
  $\hat g(x)=\hat F(x)(1-\hat F(x))/\hat F'(x)$.  Then
$g(0)>\hat g(0)$ and, from (\ref{rho(t)}), 
which can be written in the form
$g'(x)=1-F(x)-\rho(x)$, and (\ref{ineq}),
 $g'(x)-\hat g'(x)=\hat F(x)- F(x)>0$ for $0<x<y$, so that $g(y)>\hat g(y)$
and hence $F'(y)<\hat F'(y)$, which is inconsistent with (\ref{ineq}). 

\subsection{Piecewise constant profiles and extensions}

In this section we prove existence of a solution $F_\rho(x)$ of
(\ref{rho(t)}) and (\ref{condition5}), and show that this solution maximizes
$\G(\{\rho\},\{F\})$, given by (\ref{defG}), 
for a piecewise constant density profile
 \begin{equation}
 \rho(x) = r_i \ \ \ \ {\rm for} \ \ \  x_{i-1} < x < x_i, 
    \label{ri}
 \end{equation}
where $0=x_0<x_1<\cdots x_n=1$.  We will write $y_i=x_i-x_{i-1}$.  It follows
from (\ref{rho(t)}) that $F_\rho(x)$ must satisfy
 \begin{equation}
 F_\rho'(x) = A_i\,\psi_i(F_\rho(x))
     \label{F}
 \end{equation}
  for $x_{i-1}<x<x_i$, where $A_i$ is a (negative) constant and 
 \begin{equation}
  \psi_i(F) = \left( F \over r_i \right)^{r_i}
   \  \left( 1 - F \over 1 - r_i \right)^{1- r_i}.
 \label{psi}
 \end{equation}
 Continuity of $F_\rho(x)$ and $F_\rho'(x)$ implies that
$F_\rho(x_1),\ldots,F_\rho(x_{n-1})$ and the constants $A_1,\ldots,A_n$ must
satisfy
 \begin{equation}
  A_i = {1 \over y_i} \int_{F_\rho(x_{i-1})}^{F_\rho(x_i)}
   {dz \over \psi_i(z)}
  \label{Ai}
 \end{equation}
 and
 \begin{equation}
  A_i  \psi_i(F_\rho(x_i)) = A_{i+1} \psi_{i+1} (F_\rho(x_i))\;.
 \label{condition17}
 \end{equation}
 Conversely, the existence of values $F_\rho(x_1),\ldots,F_\rho(x_{n-1})$
and $A_1,\ldots,A_n$ satisfying (\ref{Ai}) and (\ref{condition17}) implies the
existence of a solution $F_\rho$ of (\ref{rho(t)}) and (\ref{condition5}),
obtained by solving (\ref{F}) on each interval $[x_{i-1},x_i]$. 

 So we need to prove that (\ref{Ai}) and (\ref{condition17}) can be solved. 
Now it follows from (\ref{F}) that
 \begin{equation}
 {\cal G}(\{\rho\},\{F\})
  = {\cal G}_0(\{r\},(F(x_0),F(x_1),\ldots,F(x_n))),
  \label{newcurlyF}
 \end{equation}
  where $\{r\}=(r_1,\ldots,r_n)$ and for $\{H\}=(H_0,\ldots,H_n)$ a
sequence satisfying
 \begin{equation}
 \rho_0=H_0\ge H_1\ge\cdots\ge H_n=\rho_1,\label{domain}
 \end{equation}
 we have defined
 \begin{equation}
{\cal G}_0(\{r\},\{H\})=
  \sum_{i=1}^n y_i \log\left( -{1 \over y_i} 
    \int_{H_{i-1}}^{H_i} {dz \over \psi_i(z)}\right) 
  -  \log(\rho_0 - \rho_1). \label{G0}
 \end{equation}
 Equation (\ref{GF}) suggests that we consider the problem of maximizing
$\G_0$. Since ${\cal G}_0$ is continuously differentiable in $\{H\}$ on the
interior of the compact domain (\ref{domain}) and equal to $-\infty$ on its
boundary, it achieves a maximum at some interior point
$\{H^*\}$ at which
 \begin{equation}
  {\partial{\cal G}_0\over\partial H_i}(\{r\},\{H^*\})
  \equiv {1\over A_{i+1} \psi_{i+1} (H_i^*)}-{1\over A_i\psi_i(H_i^*)}
  = 0, \label{crit}
 \end{equation}
 for $i=1,\ldots,n-1$, where
 \begin{equation}
  A_i = {1 \over y_i} \int_{H^*_{i-1}}^{H^*_i} {dz \over \psi_i(z)}.
  \label{Ai2}
 \end{equation}
 Since (\ref{crit}) and (\ref{Ai2}) correspond to (\ref{Ai}) and
(\ref{condition17}), we obtain a solution of the latter equations, and hence a
solution $F_\rho$ of (\ref{rho(t)}) and (\ref{condition5}) by taking
$F_\rho(x_i)=H_i^*$.  Note that the argument above could be used to construct
a solution of (\ref{rho(t)}) and (\ref{condition5}) from any
point $\{H\}$ at which 
 \begin{equation}
  {\partial{\cal G}_0\over\partial H_i}(\{r\},\{H\})
  = 0, \qquad i=1,\ldots,n-1;\label{crit2}
 \end{equation}
 since the solution is unique (see subsection 5.1), 
$\{H^*\}$ is the only point satisfying (\ref{crit2}). 

It is now easy to verify (\ref{GF}) for $\rho(x)$.  For if $F(x)$ is any
continuously differentiable monotone function with $F(0)=\rho_0$ and
$F(1)=\rho_1$, then from Jensen's
inequality applied on each interval $[x_{i-1},x_i]$,
 \begin{eqnarray}
 {\cal G}(\{\rho\},\{F\})
   &\le&  \sum_{i=1}^n y_i \log\left( -{1 \over y_i} 
    \int_{F(x_{i-1})}^{F(x_i)} {dz \over \psi_i(z)}\right) 
  -  \log(\rho_0 - \rho_1) \nonumber\\
  &\le& \sup_{H}{\cal G}_0(\{r\},\{H\}) \nonumber\\
  &=&\G_0(\{r\},\{H^*\})
   =\G(\{\rho\},\{F_\rho\}) =\F(\{\rho\}).
 \label{supG}
 \end{eqnarray}

We now discuss briefly the extension of these results to arbitrary profiles.  
A general proof of existence of a solution of (\ref{rho(t)}) 
and (\ref{condition5}) 
may be given which is independent of
the arguments above: if $\rho(x)$ is continuous then Theorem~XII.5.1 of
\cite{Hartman} implies immediately the existence of a solution $F_\rho$ of
(\ref{rho(t)}) and (\ref{condition5}), and for measurable $\rho(x)$ only
slight modifications of the proof in \cite{Hartman} are necessary.  The
uniqueness theorem for the initial value problem for (\ref{rho(t)}) implies
that the derivative of the solution must be everywhere nonzero, so that the
solution is monotonic.  Equation (\ref{GF}) may be verified for arbitrary
$\rho(x)$ by a limiting argument from the same result (proved above) for
piecewise constant densities; the key idea is 
to show that $F_\rho$ and $F'_\rho$ are, in the uniform norm, continuous
functions of $\rho$ in the $L^1$ norm.

 \subsection{Convexity of $\F(\{\rho\})$}

Finally  we show that ${\cal F}(\{\rho\})$ is convex.  Recall that $F_\rho$
is the solution of (\ref{rho(t)}) and (\ref{condition5}) corresponding to the
profile $\rho$.   From
(\ref{Euler-Lagrange}), 
 \begin{eqnarray}
{\delta{\G}\over\delta F(x)}\bigg|_{\{\rho\},\{F_\rho\}}=0,
 \end{eqnarray}
 we have 
 \begin{equation}
{\delta^2{\G}\over\delta \rho(y)\delta F(x)}\bigg|_{\{\rho\},\{F_\rho\}}
 + \int_0^1
 {\delta^2{\G}\over\delta F(u)\delta F(x)}\bigg|_{\{\rho\},\{F_\rho\}}
  {\delta F_\rho(u)\over\delta \rho(y)}(\{\rho\})\,du=0,
 \end{equation}
 and therefore 
 \begin{eqnarray}
{\delta^2{\F}\over\delta\rho(x)\delta\rho(y)}(\{\rho\})
  \;=\; {\delta\over\delta\rho(x)}\left({\delta{\cal G}\over\delta\rho(y)}
        \bigg|_{\{\rho\},\{F_\rho\}}\right) && \nonumber \\
  &&\hskip-205pt =\; 
   {\delta^2{\cal G}\over\delta\rho(x)\delta\rho(y)}\bigg|_{\{\rho\},\{F_\rho\}}
   \\
  && \hskip-185pt -\int_0^1\int_0^1
  {\delta^2{\cal G}\over\delta F(u)\delta F(w)}\bigg|_{\{\rho\},\{F_\rho\}}
     {\delta F_\rho(u)\over\delta\rho(x)}(\{\rho\})
  {\delta F_\rho(w)\over\delta\rho(y)}(\{\rho\})\,du\,dw.\nonumber
 \end{eqnarray}
 Since ${\G}(\{\rho\},\{F\})$ has a maximum at $F_\rho$ the second term
in this expression is positive semidefinite.  The first term is positive
definite:
 \begin{eqnarray}
{\delta^2{\cal G}\over\delta \rho(x)\delta \rho(y)}\bigg|_{\{\rho\},\{F_\rho\}}
  = {\delta(x-y)\over \rho(x)(1-\rho(x))}.
 \end{eqnarray}

\section{Optimal profiles\label{optprof}}

The convexity of $\F(\{\rho\})$ established above implies the existence of a
unique global minimum, corresponding to the optimal or most likely profile
$\bar\rho(x)$.  From (\ref{vd}) it follows that $\bar\rho(x)$ must satisfy
 \begin{equation}
\log\left[\bar\rho(x)(1-F(x))\over(1-\bar\rho(x))F(x)\right] = 0,
 \label{globalmin}
 \end{equation}
  where $F(x)$ is the solution of (\ref{rho(t)}) and (\ref{condition5})
corresponding to $\bar\rho(x)$.  Equation (\ref{globalmin}) leads
immediately to
 \begin{equation}
 F(x)=\bar\rho(x), 
 \end{equation}
 and  with (\ref{rho(t)}) this implies
that $F''(x)=0$; the boundary conditions (\ref{condition5}) then yield  
 \begin{equation}
 F(x) = \bar\rho(x) = \rho_0 + (\rho_1 - \rho_0) x,
 \end{equation}
 verifying that the optimal profile is as given in (\ref{rhobar}).

We may also ask for the most likely profile $\rho(x)$ under a
constraint of the form
 \begin{equation}
 \int_0^1 \psi(x) \rho(x)\, dx = K, \label{constraint}
 \end{equation}
  with fixed  weighting function $\psi(x)$ and constant $K$. 
From (\ref{vd}) one sees that $\rho(x)$ must satisfy
 \begin{equation}
{\rho(x)(1-F(x))\over(1-\rho(x))F(x)} = \exp[c\psi(x)] \label{varpsi}
 \end{equation}
 for some constant $c$, where again $F(x)$ is the solution
of (\ref{rho(t)}) and (\ref{condition5}) corresponding to $\rho(x)$.

In particular, we may impose the constraint of a fixed mean density $\rho^*$
by taking $\psi(x)\equiv1$, $K=\rho^*$.  Then from (\ref{varpsi}),
 \begin{equation}
 {\rho(x)(1-F(x))\over(1-\rho(x))F(x)} = e^c, \label{varconst}
 \end{equation}
 and we find from (\ref{rho(t)}) that $F$ must satisfy
 \begin{equation}
 {F''(x) \over F'^2(x)} = {e^c-1 \over 1 + (e^c-1)F(x)}. \label{simpleode}
 \end{equation}
 The solutions of (\ref{simpleode}) have the form $F(x)=A+B\exp(-\theta x)$,
which with the boundary conditions (\ref{condition5}) leads to
 \begin{equation}
  F(x)= {(\rho_1 - \rho_0 e^{-\theta}) - (\rho_1 - \rho_0 ) e^{-\theta x} 
    \over (1 - e^{-\theta})  };
\label{Foptimal}
 \end{equation}
 from (\ref{rho(t)}) one then finds an exponential profile
 \begin{equation}
\rho(x)= {[(1-\rho_0) e^{-\theta} - (1- \rho_1) ] 
   [ (\rho_1 - \rho_0 e^{-\theta}) e^{\theta x} - (\rho_1 - \rho_0) ] 
  \over (1 - e^{-\theta}) (\rho_1 - \rho_0) }.
\label{rhooptimal}
 \end{equation}
 Here $\theta$ is a free constant which must be chosen to satisfy the
constraint of mean density $\rho^*$.  If $\rho^*=(\rho_0+\rho_1)/2$, 
the
expected (and typical) total density in the stationary state,  then
$\theta=0$ and (\ref{rhooptimal}) reproduces the optimal profile
$\bar\rho(x)$. 
 
{\bf Remark:} (a) Unless $\theta=0$ or $\rho_0=1$, 
 \begin{equation}
 \rho(0)\equiv\lim_{x\searrow0}\rho(x) \neq \rho_0.
 \end{equation}
 Thus when we constrain the total number of particles to be different from its
most likely value, the optimal profile is discontinuous at $x=0$ (unless
$\rho_0=1$).  Similar conclusions hold at $x=1$, where a discontinuity occurs
unless $\theta=0$ or $\rho_1=0$, and,  by symmetry, in the case
$\rho_0<\rho_1$. 

(b) In the limit $\rho_1 \nearrow \rho_0$, one should scale 
$\theta\sim \rho_0 - \rho_1$
in order to obtain a meaningful answer.  The resulting optimal profile is 
constant, with an arbitrary density
$\rho_0 +\theta \rho_0 (1- \rho_0) / (\rho_0 - \rho_1)$. 

 (c) If $\theta$ is chosen such that
 \begin{equation}
 e^{-\theta} = {1 - \rho_1 + \rho_1 \fp \over 1 - \rho_0 + \rho_0 \fp},
 \end{equation}
 one can recover (\ref{singlerho}) and (\ref{singlepro}); (\ref{singlerho})
by calculating $\int_0^1\rho(x)\,dx$ from (\ref{rhooptimal}), and
(\ref{singlepro}) from (\ref{simple_expression}), (\ref{Foptimal}) and
(\ref{rhooptimal}).

 Finally, we may also impose simultaneously several constraints of the form
(\ref{constraint}).  For example, we can decompose the system as the union
of disjoint boxes and then fix the mean density in some of these, imposing
no constraints in the remainder.  Then (\ref{globalmin}) will hold in the
unconstrained boxes, so that $F(x)=\rho(x)$ will be linear there, and
(\ref{varconst}) will hold in the constrained boxes (with a box-dependent
constant $c$), so that $F$ and $\rho$ will be exponential there; the
specified mean densities in the constrained boxes and the requirement of
continuity of $F$ and $F'$ at the box boundaries then completely determines
$F$ and hence $\rho$, which will, in general, be discontinuous at the
boundaries.

Suppose, for example, that we  require that the density vanish for 
$0 < x < x_0$, with no additional constraint. 
Then
 \begin{equation}
 F(x)= \left\{
 \begin{array}{ll}
  1 - (1 - \rho_0) e^{B x},&\hbox{if $ x \leq x_0$,}\\
 \rho_1 - (1 - \rho_0)  B e^{B x_0} (x-1), &\hbox{if $ x > x_0$,}
 \end{array}\right.
 \end{equation}
 and  continuity of $F$  at $x=x_0$  requires that $B$ satisfy
 \begin{equation}
{1 - \rho_1 \over 1 - \rho_0} = [1 + B (1 - x_0 )] e^{B x_0}.
\label{CONDITION}
 \end{equation}
 From (\ref{simple_expression}) we find that $\F(\{\rho\})$, which in this
case is simply the probability that $\rho(x)$ vanish for $0<x<x_0$, is
given by 
 \begin{equation}
  \F(\{\rho\})  =   -\log\left( \rho_0 - \rho_1 \over B \right)
  + (1 - x_0) \log\left(  1 - \rho_1 \over 1 + B (1 - x_0) \right).
 \end{equation}
 If now we take $\rho_0 \to 1$ with $\rho_1$ and $x_0$
fixed,  (\ref{CONDITION}) implies that
$B \simeq -  x_0^{-1} \log (1 - \rho_0) $ and therefore
 \begin{equation}
 \F(\{\rho\})  \simeq   x_0 \log( - \log(1 - \rho_0)) \;.
 \end{equation}
 Thus the probability of finding zero density in the box $x < x_0$
when $\rho_0=1$ is super-exponentially small.

\section{Small fluctuations in the profile\label{fluctuations}}

We have already given in (\ref{fluct}) the formula for the probability of
small fluctuations in the total number of particles in the system.  In this
section we compute, directly from the large deviation functional
$\F(\{\rho\})$, the covariance of small fluctuations in the profile
around the stationary
profile $\rhob(x)=(1-x)\rho_0+x\rho_1$ (\ref{rhobar}).  
 We show also that the
result agrees with with a direct computation 
of the correlation functions  from the microscopic measure
describing the SNS  (note that the possibility that 
these two computations might
disagree had been conjectured in \cite{E}).

\subsection{Small fluctuations from large deviations}

For  a small fluctuation of the density profile around its optimum $\rhob(x)$,
 \begin{equation}
 \rho(x) = \rhob(x) + \epsilon(x) \;,\qquad  \epsilon(x)<<1\;,
 \end{equation}
 there will be a corresponding variation of $F(x)$,
 \begin{equation}
 F(x) = \rhob(x) + \phi(x).
 \end{equation}
 From (\ref{condition5}) we have  $\phi(0)=\phi(1)=0$, and from 
(\ref{rho(t)}) $\phi(x)$ will satisfy, to first order in $\epsilon(x)$,
 \begin{equation}
 \epsilon(x) = \phi(x) + { \rhob(x) (1 - \rhob(x)) \over \rhob'(x)^2}
 \phi''(x).
\label{eps}
 \end{equation}
 From (\ref{simple_expression}), again to lowest (quadratic) order in
$\epsilon(x)$,
\begin{eqnarray}
\F(\{\rho\})
 = - {1 \over 2} \int_0^1 dx \left\{  {\phi'(x)^2 \over \rhob'^2} - 
 { \rhob(x) (1 - \rhob(x)) \over \rhob'^4} \phi''(x)^2
   \right\}.
\label{quadratic}
\end{eqnarray}

One can rewrite (\ref{eps}) as
 \begin{equation}
\epsilon(x) = \int_0^1 dy \  C(x,y) \  \phi''(y)\;,
\label{eps1}
 \end{equation}
 with $C(x,y)$  given by
 \begin{equation}
   C(x,y)  =-(1-x) y \theta(x-y) - x (1-y) \theta(y-x) 
   +{\rhob(x) (1-\rhob(x) ) \over \rhob'^2} \delta(y-x),\label{Cexplicit}
 \end{equation}
 where $\theta(x)$ is the Heaviside function,
or with a more compact notation,
 \begin{equation}
 C(x,y)=\Delta^{-1}(x,y)+{\rhob(x) (1-\rhob(x) ) \over \rhob'^2} \delta(x-y).
  \label{C}
 \end{equation}
 Here  $\Delta$ is the second derivative operator with Dirichlet boundary 
conditions at $x=0$ and $x=1$.
The expression (\ref{quadratic}) can then be written as
\begin{eqnarray}
 \F(\{\rho\})
 &=&  {1 \over 2 \rhob'^2} \int_0^1 dx  \int_0^1 dy  
   \  \phi''(x) \  C(x,y) \  \phi''(y)\nonumber\\
 &=& - {1 \over 2 \rhob'^2} \int_0^1 dx  \int_0^1 dy \   
   \epsilon(x) \  C^{-1}(x,y) \  \epsilon(y)\;,
\label{quadratic1}
\end{eqnarray}
 where we have used (\ref{eps1}).  This formula expresses the fact that,
for large $N$, density fluctuations are approximately Gaussian with
covariance matrix $\rhob'^2C(x,y)/N$.  Using the explicit expression
(\ref{Cexplicit}) of $C$, this leads to the formula for the
correlations $\langle \epsilon(x) \epsilon(y) \rangle$:
 \begin{eqnarray}
 \langle \epsilon(x) \epsilon(y) \rangle 
  &=& {\rhob'^2 \  C(x,y) \over N} \nonumber\\
  &=& {1\over N} \Bigl\{- \rhob'^2\bigl[(1-x) y \theta(x-y) 
   +  (1-y)x \theta(y-x)\bigr] \nonumber \\
   && \hskip40pt  +\;\rhob(x) (1-\rhob(x))  \delta(x-y) \Bigr\}\;.
\label{smallfluct}
 \end{eqnarray}
This result was obtained earlier by Spohn \cite{HS1},  and shown there
to agree with the results from fluctuating hydrodynamics.

\subsection{Microscopic correlation functions}

One may also compute correlations directly from 
from the algebra (\ref{alg1}-\ref{alg3}) and the normalization factor
(\ref{Z0}).  By recursions over $i$ and $j$ one finds 
for $1 \leq  i \leq N$ that 
 \begin{eqnarray}
   \langle W |(D+E)^{i-1} D (D+E)^{N-i} | V \rangle &&\nonumber\\ 
  &&\hskip-150pt = \rho_0   \langle W | (D+E)^N | V \rangle  
-(a+i-1)  \langle W | (D+E)^{N-1} | V \rangle \;, \label{single}
 \end{eqnarray}
 and for $1 \leq j < i \leq N$ that
 \begin{eqnarray}
   \langle W |(D+E)^{j-1} D (D+E)^{i-j-1} D (D+E)^{N-i} | V \rangle
    \nonumber\\ 
   &&\hskip-190pt   = \rho_0^2   \langle W | (D+E)^N | V \rangle  
    \nonumber \\ 
   &&\hskip-160pt   \nonumber
    -\rho_0 (2 a+i+ j -2)  \langle W | (D+E)^{N-1} | V \rangle  \\
      &&\hskip-160pt 
   +  ( a+ j -1)(a+i-2)  \langle W | (D+E)^{N-2} | V \rangle  \;.
  \label{double}
 \end{eqnarray}
 Using (\ref{Z0}), one obtains from (\ref{single}) the average profile
 \begin{equation}
\langle \tau_i \rangle = \rho_0 + {a+i-1 \over a+b + N-1} (\rho_1 - \rho_0)\;,
\label{tau}
 \end{equation}
 and from (\ref{double}) the truncated (microscopic) correlation function:
for $j < i $,
 \begin{equation}
\langle \tau_j \tau_i \rangle 
-\langle \tau_j  \rangle \langle \tau_i \rangle =  - (\rho_1 - \rho_0)^2
 {(a+j-1) ( b+ N -i)  \over (a+b + N-1)^2 (a+b+N-2)} \;.
\label{tautau}
 \end{equation}
 Equations (\ref{tau}) and (\ref{tautau}) agree with the corresponding
formulas in \cite{HS1}, where for rates satisfying
$\alpha + \beta = a = \gamma + \delta = b = 1$, random walk methods were
used to calculate the one particle and pair correlation for this system.

In the large $N$ limit, (\ref{tau}) yields the linear profile
(\ref{rhobar}), and by summing (\ref{tautau}) over $i$ and $j$, one
recovers (\ref{variance}) 
Moreover, we may recover  (\ref{smallfluct}): if we  divide the system into
boxes of size $\Delta x$ (with $1\ll \Delta x^{-1}\ll N$)
and write, for $x$ and $y$  multiples of $\Delta x$, 
 \begin{equation}
 \epsilon(x)\simeq
   {1\over N\Delta x}\sum_{i=Nx}^{N(x+\Delta x)}\tau_i
   -  \rho(x) ,\qquad
 \epsilon(y)\simeq
   {1\over N\Delta x}\sum_{i=Ny}^{N(y+\Delta x)}\tau_i
  -  \rho(y) ,
 \end{equation}
 then 
 \begin{eqnarray}
\langle \epsilon(x) \epsilon(y) \rangle 
- \langle \epsilon(x) \rangle \langle \epsilon(y) \rangle 
  &\simeq& {1 \over  N^2\Delta x^2} \left\{
\delta_{x,y} \sum_{i=Nx}^{N(x+\Delta x)}
(\langle  \tau_i \rangle -
 \langle \tau_j \rangle^2 )\right.\nonumber\\
 &&\hskip-30pt + \left.
\sum_{j=Nx}^{N(x+\Delta x)}
\sum_{i=Ny}^{N(y+\Delta x)} (\langle \tau_j \tau_i \rangle -
 \langle \tau_j \rangle 
 \langle \tau_i \rangle )  \right\} ,
 \end{eqnarray}
 and this, with (\ref{tau}) and (\ref{tautau}), yields (\ref{smallfluct}).

\section{Examples\label{examples}}

Although in general the calculation of the large deviation function
$\F(\{\rho\})$ for a given profile $\rho(x)$ is not easy, since it requires
the solution of the nonlinear second order differential equation
(\ref{rho(t)}), there are nevertheless some cases for which it can be done.
Piecewise constant profiles were discussed in Section~\ref{continuum}.2; in
particular, for a profile of constant density $r$ the large deviation
functional is given by (\ref{constantrho}), and this function can be found
explicitly when $r=0$ or $1$ (in which case $F(x)$ is exponential) or
$r=1/2$ (when $F(x)$ is sinusoidal). $\F(\{\rho\})$ can also be found
explicitly in some cases for the optimal profiles under constraint,
discussed in Section~\ref{optprof}. 

One may also construct examples by specifying $F(x)$ and obtaining
$\rho(x)$ from (\ref{rho(t)}) and $\F(\{\rho\})$ from
(\ref{simple_expression}). One must check in each case that
$0\le\rho(x)\le1$, since this is not guaranteed by (\ref{rho(t)}).
In the
remainder of this section we give two examples of this type which address
the question: for what profiles $\rho(x)$ is $\F(\{\rho\})$ infinite?  
 If $1>\rho_0>\rho_1>0$ then $\F(\{\rho\})<\infty$;
this follows from (\ref{simple_expression}) and the fact that,
by the
uniqueness theorem for the initial value problem for (\ref{rho(t)}) (see
remark at the end of Section~5.2), $F'(x)$ cannot vanish.
Suppose then that $\rho_0=1$ (analysis of the case
$\rho_1=0$ is similar).  The examples below suggest that
$\F(\{\rho\})=\infty$ when $\lim_{x\to0}\rho(x)=0$ and this limit is
approached faster than any power of $x$.  This is also 
supported by the example at the end of Section~\ref{optprof}, which 
shows that $\F(\{\rho\})=\infty$ if $\rho_0=1$ and $\rho(x)$ vanishes
identically on an arbitrarily small interval $0<x<x_0 $.  However, we
have not formulated any sharp conjecture.

In both examples we take $\rho_0=1$. For the first, define
 \begin{equation}
 F(x)=1-(1-\rho_1)e^{c(1-1/x^n)}, \label{Fexamp1}
 \end{equation}
  with $c$ a positive constant. Note that $F$ is monotone decreasing and
satisfies the boundary conditions (\ref{condition5}). Then 
 \begin{eqnarray}
   F'(x) &=& -{nc\over x^{n+1}}(1-F(x)),\label{Fprime1}\\
   F''(x) &=& -\left(nc\over x^{n+1}\right)^2\left(1-{n+1\over cn}x^n\right)
   (1-F(x)),\label{Fdprime1}
 \end{eqnarray}
 and hence  from (\ref{rho(t)}),
 \begin{equation}
 \rho(x) ={n+1\over cn}x^nF(x) . \label{rho1}
 \end{equation}
 Thus $\rho(x)$ vanishes like $x^n$ at $x=0$, while
$\lim_{x\to1}\rho(x)=\rho_1(n+1)/cn$.  Certainly we may choose $c$ (e.g.,
$c=(n+1)/n$) to ensure that $0\le\rho(x)\le1$ for all $x$ (note that by
choosing $c$ sufficiently small we see that (\ref{rho(t)}) for an admissible
$F$ does not guarantee $0\le\rho(x)\le1$).  If we then write
(\ref{simple_expression}) in the form
 \begin{eqnarray}
\F(\{\rho\}) 
 &=&\int_0^1\,dx\left\{\rho(x)\log{\rho(x)(1-F(x))\over(1-\rho(x))F(x)}
    +\log{(1-\rho(x))\over(1-\rho_1)}\right.\nonumber\\
 &&\hskip50pt + \left.\log{-F'(x)\over(1-F(x))}\right\} \label{LD1}
 \end{eqnarray}
 then (\ref{Fexamp1}), (\ref{Fprime1}), and (\ref{rho1}) show that
$\F(\{\rho\})<\infty$. 

Finally, we again take  $\rho_0=1$ and $c>0$ and define
 \begin{equation}
 F(x)=1-(1-\rho_1)e^{c(e-e^{1/x})}\;. \label{Fexamp2}
 \end{equation}
  Then 
 \begin{eqnarray}
   F'(x) &=& -{ce^{1/x}\over x^2}(1-F(x)),\label{Fprime2}\\
   F''(x) &=& -\left(ce^{1/x}\over x^2\right)^2\left(1-{2x+1\over c}e^{-1/x}\right)
   (1-F(x)),\label{Fdprime2}
 \end{eqnarray}
 and 
 \begin{equation}
 \rho(x) ={2x+1\over c}e^{-1/x}F(x) . \label{rho2}
 \end{equation}
 Now $\rho(x)$ vanishes faster than any  power of $x$ at $x=0$. 
Again we may choose $c$ (e.g.,
$c=3$) so that $0\le\rho(x)\le1$ for all $x$; if we then write
(\ref{simple_expression}) as in (\ref{LD1}) 
we see from (\ref{Fexamp2}), (\ref{Fprime2}), and (\ref{rho2}) that
although the first two terms are finite, the last contributes
$\int_0^1\,dx/x$, so that $\F(\{\rho\})=\infty$.

\section{Conclusion\label{conclusion}}

In the present work, we have seen that the large deviation functional 
$\F(\{\rho\})$ of the
density profile  $\rho(x)$ can be calculated for the SSEP,
 starting from the known
weights of the microscopic configurations.  We find that the large deviation
function is in general finite,  although there are counterexamples
(see (\ref{constantrho}) and section 
\ref{examples}).  

The simplest way we found to write our results  is a parametric form 
(\ref{simple_expression}--\ref{condition5})  
in which both the large deviation function and the density profile 
are expressed in terms of
a monotonic function $F(x)$ varying between $\rho_0$ and $\rho_1$, 
the densities of the two reservoirs at the two ends of the system.

An interesting question posed by the present work is how the additivity
property of the free energy of equilibrium systems is modified in the
nonequilibrium case.  We first note that if a system of $N$ sites, in contact
with left and right reservoirs at chemical potentials corresponding to
densities $\rho_a$ and $\rho_b$, is described by a macroscopic coordinate $x$
satisfying $a\le x\le b$ (rather than $0\le x\le 1$), then the probability
$P$ of observing a profile $\rho(x)$, $a\le x\le b$, satisfies
\begin{equation}
\log P \simeq -{N\over b-a} {\cal F}_{[a,b]}(\{\rho\}; \rho_a, \rho_
b),
\label{pab}
\end{equation}
 where 
 \begin{eqnarray}
{\cal F}_{[a,b]}(\{\rho\};\rho_a,\rho_b)
   &\equiv& \int_a^b dx
   \left\{ \rho(x) \log \left({ \rho(x) \over F(x)}\right)
    \right.
     \nonumber\\
 && \hskip-70pt
  + \left. (1- \rho(x)) 
  \log \left( 1 - \rho(x) \over 1 -F(x) \right)
  + \log \left({ (b-a) F'(x) \over \rho_b - \rho_a} \right) \right\}\;,
\label{simple_expression_ab}
\end{eqnarray}
 and $F(x)$ is related to $\rho(x)$ 
on the   interval $a<x <b$ by (\ref{rho(t)}) 
 with the boundary conditions
 \begin{equation}
F(a)= \rho_a \;,\ \ \ \  F(b)=\rho_b \; .
 \end{equation}
(Note that $F'(x)$ has the same sign as $\rho_b - \rho_a$,
so that the argument of the $\log$ is positive  in the last term of
(\ref{simple_expression_ab}).)

Now consider two systems, of $Nu$ and $N(1-u)$ sites respectively (with
$0<u<1$), with the left system in contact with left and right reservoirs at
chemical potentials corresponding to densities $\rho_0$ and $\rho_m$ and
the right system with reservoirs corresponding to densities $\rho_m$ and
$\rho_1$ (with no other relation between them, in particular with no
particles jumping directly from one system to the other).  If $\rho(x)$ is
a profile on $0<x<1$ and $\rho^{(1)}(x)$ and $\rho^{(2)}(x)$ are its
restrictions to the intervals $0<x<u$ and $u<x<1$, then we would like to
relate the probability $P_N$ of observing $\rho(x)$ to the probabilities
$P^{(1)}_N$ and $P^{(2)}_N$ of observing $\rho^{(1)}(x)$ and $\rho^{(2)}(x)$ in
the two subsystems.  If $\F$ were truly local we would have
$P_N\simeq P^{(1)}_NP^{(2)}_N$.   
The most naive generalization of this idea here
would be that
$P_N\simeq \sup_{\rho_m}(P^{(1)}_NP^{(2)}_N)$ or, since  from
(\ref{pab})
$\log P^{(1)}_N\simeq-N{\F}_{[0,u]}(\{\rho^{(1)}\};\rho_0,\rho_m)$ and
$\log P^{(2)}_N\simeq-N{\F}_{[u,1]}(\{\rho^{(2)}\};\rho_m,\rho_1)$,
equivalently that  
${\F}_{[0,1]}(\{\rho\};\rho_0,\rho_1) 
  =\inf_{\rho_m} \bigl[{\F}_{[0,u]}(\{\rho^{(1)}\};\rho_0,\rho_m) 
  +{\F}_{[u,1]}(\{\rho^{(2)}\};\rho_m,\rho_1)\bigr]$.  One
can check, however, that this is not true.

Instead, we observe that if we define a ``modified free energy'' ${\cal H}$ by
 \begin{equation}
  {\cal H}_{[a,b]}(\{\rho\};\rho_a,\rho_b) 
   = \F_{[a,b]}(\{\rho\};\rho_a,\rho_b)  
   +  (b-a)\log\left(\rho_a - \rho_b\over b-a\right)\;,
\label{Hdefab}
\end{equation}
 then we obtain from (\ref{GF}) the ``additivity'' property
\begin{equation}
{\cal H}_{[0,1]}(\{\rho\};\rho_0,\rho_1) 
  = \sup_{\rho_m}\bigl\{{\cal H}_{[0,u]}(\{\rho^{(1)}\};\rho_0,\rho_m) 
    + {\cal H}_{[u,1]}(\{\rho^{(2)}\};\rho_m,\rho_1) \bigr\}.
\label{additionab}
\end{equation}
  The occurrence in (\ref{additionab}) of the supremum rather than infimum is a
   consequence of (\ref{simple_expression}), but we do not understand its physical basis at
   this time.

It is perhaps surprising that the additivity property (\ref{additionab})
suffices to determine the large deviation functional completely.  For suppose
we divide our system as above, but into $n$ rather than 2 subsystems, with
division points $0=x_0<x_1<\cdots<x_n=1$, and denote the corresponding
reservoir densities by $\rho_0=F_0>F_1>\cdots>F_{n-1}>F_n=\rho_1$.  Then
iterating (\ref{additionab}) we find that
 \begin{eqnarray}
\F_{[0,1]}(\{\rho\};\rho_0,\rho_1) &&\nonumber\\
  &&\hskip-100pt=\;\sup_{F_1\,\ldots,F_{n-1}}
  \sum_{j=1}^n \left\{\F_{[x_{j-1},x_j]}(\{\rho\};F_{j-1},F_j) 
    + y_j\log\left(F_{j-1}-F_j\over
  (\rho_0-\rho_1)y_j\right)\right\}, \label{bigsup}
 \end{eqnarray}
  where $y_j=x_j-x_{j-1}$.  When $n$ is large and all $y_j$ are small, 
 \begin{equation}
F_j-F_{j-1}\simeq y_jF'(x_j);\label{simpleapprox}
 \end{equation}
 moreover, since $F_j\simeq F_{j-1}$, that is, the two reservoirs at the ends
of the interval $[x_{j-1},x_j]$ have essentially the same chemical
potentials, we can expect each this subsystem to be in equilibrium, so that
from (\ref{equil}),
 \begin{eqnarray}
  \F_{[x_{j-1},x_j]}(\{\rho\};F_{j-1},F_j)&&\nonumber\\
 &&\hskip-100pt \simeq y_j\left[ \rho(x_j) \log\left(\rho(x_j)\over F(x_j)\right)
   + (1- \rho(x_j)) \log\left(1-\rho(x_j)\over 1-F(x_j)\right)\right]. \label{FsimFeq}
 \end{eqnarray}
 By substituting (\ref{simpleapprox}) and (\ref{FsimFeq}) into (\ref{bigsup})
and taking the limit $n\to\infty$ with $y_j\to0$ for all $j$, we obtain our
basic result (\ref{GF}). 

Clearly it would be of great interest to give a direct derivation of the
additivity property (\ref{additionab}), and to know if this property is
limited to the SSEP or whether similar relations hold for more general
systems.

\section*{Acknowledgments:}

 We thank L.~Bertini, A.~De~Sole, D.~Gabrielle, G.~Giacomin, 
G.~Jona-Lasinio, C.~Landim, E.~Lieb, J.~Mallet-Paret,
R.~Nussbaum, E.~Presutti, and R.~Varadhan for very helpful discussions and
communications. 
The work of J.~L.~Lebowitz was supported by NSF Grant DMR--9813268, AFOSR
Grant F49620/0154, DIMACS and its supporting agencies, the NSF under
contract STC-91-19999 and the N.  J.  Commission on Science and
Technology, and NATO Grant
PST.CLG.976552.  J.L.L. and E.~R.~Speer acknowledge the hospitality of the
I.H.E.S. in the spring of 2000, where this work was begun.

\appendix

\section{Derivation of (\ref{Zresult}) and (\ref{gdef})
  \label{algebra}}

\subsection{A first  consequence of (\ref{alg1})}

Let us first prove  that  if $D$ and $E$ are operators satisfying
(\ref{alg1}),
 \begin{equation}
 DE- ED= D+E\;,\label{alg1again}
 \end{equation}
 then
 \begin{equation}
 e^{x D + y E} = \left( (x-y) e^y \over x e^y - y e^x \right)^E
\   \left( (x-y) e^x \over x e^y - y e^x \right)^D\;.
\label{identity1}
 \end{equation}
 Equation (\ref{identity1}) and similar equations below are 
to be interpreted in terms of formal 
power series in $x$ and $y$.  

One can easily check from (\ref{alg1again}) that for all $p \geq 0$ 
 \begin{equation}
 D E^p = (E+1)^p D + E (E+1)^p - E^{p+1} \;,
 \end{equation}
 which means that for  "arbitrary functions" $f(E)$,
one has
 \begin{equation}
 D f(E) = f(E+1) D + E [ f(E+1) - f(E)] \;.
 \end{equation}
 Then if one tries to write $ e^{z (x D + y E)}$ under the form 
 \begin{equation}
  e^{z (x D + y E)} =
 e^{t E} e^{uD}\;,
 \end{equation}
 one gets  that
 \begin{equation}
 {dt \over dz} = x (e^t -1) + y \;,
 \end{equation}
 \begin{equation}
 {du \over dz} = x e^t  \;,
 \end{equation}
 and  by integrating over $z$, one obtains
 (\ref{identity1}).

\subsection{Other consequences of (\ref{alg1})}
Using (\ref{identity1}), one derives easily the following three identities:
 \begin{equation}
 e^{x D + y E} = \left( x e^x  - y e^y \over (x-y) e^y \right)^D
 \  \left( x e^x  - y e^y \over (x-y) e^x \right)^E\;,
\label{identity2}
 \end{equation}
 \begin{equation}
e^{x D} e^{y E} = \left( e^y \over e^x + e^y - e^{x+y} \right)^E
\  \left( e^x \over e^x + e^y - e^{x+y} \right)^D \;,
\label{identity3}
 \end{equation}
 \begin{equation}
e^{x E} e^{y D} = \left( e^x +  e^y -1 \over  e^x  \right)^D
\  \left( e^x + e^y -1 \over e^y  \right)^E \;.
\label{identity4}
 \end{equation}
Then combining (\ref{identity2}) and (\ref{identity3}), one can also show
that
\begin{eqnarray}
e^{x E}  e^{y D} &=& 
\left( (\rho_0 - \rho_1) e^x \over 
1 - (1- \rho_0) e^x  -  \rho_1 e^y  \right)^{\rho_0 E -  (1- \rho_0) D }
\nonumber\\
&&\hskip40pt\times\left( (\rho_0 - \rho_1) e^y \over 
1- (1 - \rho_0) e^x  -  \rho_1 e^y  \right)^{(1- \rho_1)  D-  \rho_1  E },
\label{identity5}
\end{eqnarray}
and that
\begin{eqnarray}
\label{recursion}
e^{u E} e^{v D} e^{xD + y E}
   &=& \left( (x-y) e^{u+y} \over (x-y) e^y + y (e^{v+y} - e^{v+x})
\right)^E
 \nonumber\\
 &&\hskip30pt\times
   \left( (x-y) e^{v+x} \over (x-y) e^y + y (e^{v+y} - e^{v+x}) \right)^D.
\end{eqnarray}

\subsection{Derivation of (\ref{Zresult})}

As a consequence of (\ref{recursion}) we see that if $\{ x_n \}$ 
and $\{ y_n \}$ are two  
sequences and if the 
sequences $\{u_n\}$ and $\{v_n\}$ are defined by the recursion
 \begin{equation}
e^{u_{n+1} E} e^{v_{n+1} D}  =
e^{u_n E} e^{v_n D} e^{x_{n+1}D + y_{n+1} E} ,
\label{recursion1}
 \end{equation}
with the initial condition
 \begin{equation}
 u_0=v_0=0,
 \end{equation}
 then the general expressions of $u_n$ and $v_n$ are given by
 \begin{equation}
e^{v_n} = \left[ e^{\sum_{i=1}^n y_i - x_i}  + \sum_{i=1}^n {y_i \over x_i - y_i} (e^{y_i-x_i} -1) e^{ \sum_{j>i} y_j-x_j} \right]^{-1}\;,
\label{un}
 \end{equation}
 \begin{equation}
e^{u_n} =  e^{\sum_{i=1}^n y_i-x_i}  \left[  e^{\sum_{i=1}^n y_i - x_i} + \sum_{i=1}^n {y_i \over x_i - y_i} (e^{y_i-x_i} -1) e^{ \sum_{j>i} y_j-x_j} \right]^{-1}\;.
\label{vn}
 \end{equation}
Therefore           
 \begin{equation}
 e^{\mu_1 \fp_1 D + \mu_1 E}   \cdots e^{\mu_k \fp_k D + \mu_k E} 
=  e^{u_n  E} \  e^{v_n D}\;,
\label{Prod}
 \end{equation}
where $u_n$ and $v_n$ are given by
 \begin{equation}
e^{v_n} = \left[ e^{\sum_{i=1}^n \mu_i (1 - \fp_i)}  + \sum_{i=1}^n {1 \over \fp_i - 1} (e^{\mu_i (1 - \fp_i)} -1) e^{ \sum_{j>i} \mu_j (1 - \fp_j)} \right]^{-1}\;,
\label{vn1}
 \end{equation}
\begin{eqnarray}
e^{u_n} &=&  e^{\sum_{i=1}^n \mu_i (1 - \fp_i)} \label{un1}\\
 &&\hskip-10pt\times \left[ e^{\sum_{i=1}^n \mu_i (1 - \fp_i)}  + \sum_{i=1}^n {1 \over
 \fp_i - 1} (e^{\mu_i (1 - \fp_i)} -1) e^{ \sum_{j>i} \mu_j (1 - \fp
_j)} \right]^{-1}.\nonumber
\end{eqnarray}
Then using (\ref{identity5}), one can rewrite (\ref{Prod}) as
 \begin{equation}
 e^{\mu_1 \fp_1 D + \mu_1 E}   \cdots e^{\mu_k \fp_k D + \mu_k E} 
=  e^{U_n [- D + \rho_0 (D + E)]  } \  e^{V_n [ D - \rho_1 (D+E) ] }\;,
\label{Prod1}
 \end{equation}
where 
 \begin{equation}
  e^{V_n}={\rho_1-\rho_0\over g}\;,\qquad 
  e^{U_n}=e^{V_n+\sum_{i=1}^n\mu_i(1-\fp_i)},\label{VnUndef}
 \end{equation}
 with $g$ given by (\ref{gdef}).
Lastly, with  $\rho_0$, $\rho_1$, $a$, and 
$b$ given by (\ref{rho12}) and (\ref{abdef}),  
the algebraic rules (\ref{alg2}) and (\ref{alg3}) can be written as
 \begin{eqnarray}
[ D - \rho_1(D +  E)] | V \rangle =   b | V \rangle \;,
\label{alg2bis} \\
 \langle W | [ - D + \rho_0 (D + E  )]  = a \langle W | \;,  
\label{alg3bis} 
 \end{eqnarray}
and one  obtains (\ref{Zresult}) from (\ref{Prod1}).

\section{Derivation of (\ref{feq}--\ref{rj})\label{asymptotics}}

Consider a system of  $N$ sites, divided into $n$ boxes, with a
fugacity $\fp_1$ on the $N_1$ first sites  on the left, 
then $\fp_2$ on the next $N_2$ sites
and so on, $\fp_n$ on the last $N_n$ sites.
Clearly we have
 \begin{equation}
 N= N_1 + N_2 + ... + N_n \;.
 \end{equation}
 Let us   define   $\Omega$ by  
\begin{eqnarray}
\Omega &=&  { \langle W | (\fp_1 D + E)^{N_1}  ...
 (\fp_n D + E)^{N_n} | V \rangle \over
 \langle W   | V \rangle } \nonumber\\
  &=&
  \sum_{0\le M_i\le N_i}
 \fp_1 ^{M_1} \cdots \fp_n ^{M_n} 
{\Omega_{N_1,\dots,N_n}(M_1,\ldots,M_n) \over \langle W | V \rangle} \;,
 \label{Omega}
\end{eqnarray}
 (see (\ref{Zdef})), and $\Omega_0$ as in (\ref{Z0}),
 \begin{equation}
\Omega_0=  { \langle W | (D + E)^{N}  
 | V \rangle \over
 \langle W   | V \rangle } \;.
\label{Omega_0}
 \end{equation}
 Here we have suppressed the dependence of $\Omega_0$ on $N$ and of $\Omega$ on
$N_1,\ldots,N_n$ and $\fp_1,\ldots,\fp_n$. 
Suppose that for large $N_1, .., N_n$, the quantity defined by (\ref{Omega})
has the following  behavior
 \begin{equation}
\Omega \sim e^{ N h(y_1,y_2,...y_n; \fp_1,...\fp_n) }  N_1 ! .... N_n ! \;,
\label{Omegaasympt}
 \end{equation}
where 
 \begin{equation}
 y_i = {N_i \over N}  \;.
 \end{equation}

Clearly one has that the average 
density $r_i$ of particles in box $i$, in the
ensemble with fugacities $\fp_1,\ldots,\fp_n$,  is
given by
 \begin{equation}
r_i y_i = {\langle M_i \rangle_{\fp_1,\ldots,\fp_n}
   \over N} = {1 \over N} {\partial \log \Omega \over \partial \log \fp_i}
  \simeq  {\partial  h \over \partial  \log \fp_i} \;.
\label{riexp}
 \end{equation}
If we assume that the distribution of the $M_i$ is strongly peaked near their
mean values then
\begin{eqnarray}
\Omega&\equiv&\Omega_0\sum_{M_1,\ldots,M_n}\fp_1^{M_1}\cdots\fp_n^{M_n}
   P_{N_1,\ldots,N_n}(M_1,...M_n) \nonumber\\
 &\approx& \Omega_0\fp_1^{r_1y_1N}\cdots\fp_n^{r_ny_nN}
    P_{N_1,\ldots,N_n}(r_1y_1N,...r_ny_nN)\;,  \label{Omega2}
\end{eqnarray}
 where $P_{N_1,...N_n}(M_1, ... M_n)$ denotes the probability 
computed in the ensemble in which $\fp_i=1$ for all $i$.  From
(\ref{Omegaasympt}) and (\ref{Omega2}),
 \begin{eqnarray}
{\log   P_{N_1,...N_n}a(M_1,...M_n) \over N} && \label{logP}\\
 &&\hskip-80pt \simeq    h(y_1,...y_n; \fp_1,...\fp_n)  - r_1 y_1  \log \fp_1  
... -  r_n y_n \log \fp_n    +K \;, \nonumber
 \end{eqnarray}
with $M_i=r_iy_i$ and $K$ given by (see (\ref{Z0}))
 \begin{equation}
K=  - {\log \Omega_0 \over N} +   \sum_{i=1}^n {\log(N_i!) \over N}
\simeq  \log( \rho_0 - \rho_1) +  \sum_{i=1}^n y_i \log y_i \;.
\label{logPbis}
 \end{equation}
We thus obtain $P_{N_1,...N_n}(M_1, ... M_n)$ in a 
parametric form (\ref{logP}): 
as we vary the $\fp_i$'s, the densities $r_i$ in the boxes vary
according to (\ref{riexp}).

Let us now see how we can extract the function $h$ which
appears in (\ref{Omegaasympt}) from (\ref{Zdef}--\ref{gdef}).
If we consider the generating function (\ref{Zdef}),
 \begin{equation}
 Z= \sum_{N_1=0}^\infty
... \sum_{N_n=0}^\infty  {\mu_1^{N_1} \over N_1!}
... {\mu_n^{N_n} \over N_n!}
 { \langle W | (\fp_1 D + E)^{N_1}  ...
 (\fp_n D + E)^{N_n} | V \rangle \over
 \langle W   | V \rangle }\;,
 \end{equation}
 and  use the asymptotic form (\ref{Omegaasympt}),
we see that $Z$ becomes singular along a manifold
 \begin{equation}
g=0\;,
\label{hypersurface}
 \end{equation}
where
 \begin{equation}
g= \max_{y_1,...y_n {\rm with} \  y_1+...y_n=1}
 [h(y_1,...y_n;\fp_1,...\fp_n) + y_1 \log \mu_1 + ... y_n \log \mu_n ] .
\label{gdefa}
 \end{equation}
Along this manifold, we have obviously
 \begin{equation}
 h = - \sum_{i=1}^n y_i \log \mu_i\;,
 \end{equation}
 so that (\ref{logP}) and (\ref{logPbis}) reduce to  (\ref{Pro}).
Moreover we see from (\ref{gdefa}) that
 \begin{equation}
 y_i = {1 \over C} {\partial g \over \partial \log \mu_i}\;,
 \end{equation}
 where the constant $C$ is a Lagrange multiplier associated to the constraint 
that $\sum_i y_i =1$, and this  establishes (\ref{xj}).
Lastly (\ref{rj}) follows simply from the fact that along the hypersurface
given by (\ref{hypersurface}) and (\ref{gdefa}), one has
 \begin{equation}
  {\partial g \over \partial \log \fp_i}=
 {\partial h \over \partial \log \fp_i}\;,
 \end{equation}
 which shows that (\ref{rj}) follows from (\ref{riexp}).

\end{document}